\documentclass[a4paper]{spie}  

\usepackage[]{graphicx}
\usepackage[]{keyval}
\usepackage[]{caption}
\usepackage[]{subfig}

\title{Mid-infrared astronomy with the E-ELT: Performance of METIS}

\author{S. Kendrew\supit{a}, L. Jolissaint\supit{a}, B. Brandl\supit{a}, R. Lenzen\supit{b}, E. Pantin\supit{c}, A. Glasse\supit{d}, J. Blommaert\supit{e}, L. Venema\supit{f}, R. Siebenmorgen\supit{g}, F. Molster\supit{h}
\skiplinehalf
\supit{a}Leiden Observatory, University of Leiden, PO Box 9513, 2300 RA Leiden, Netherlands; \\
\supit{b} Max Planck Institute for Astronomy, K\"{o}nigstuhl 17, 69117 Heidelberg, Germany;\\
\supit{c}CEA Saclay, Gif-sur-Yvette, 91191 cedex, France;\\
\supit{d}UK Astronomy Technology Centre, Royal Observatory of Edinburgh, Blackford Hill, Edingburgh EH9 3HJ, United Kingdom;\\
\supit{e}Katholieke Universiteit Leuven, Institute of Astronomy, Celestijnenlaan 200D Bus 2401, 3001 Leuven, Belgium;\\
\supit{f}Astron, Oude Hoogeveensedijk 4, 7991 PD Dwingeloo, Netherlands;\\
\supit{g}ESO, Karl SchwarzschildStrasse 2, 85748 Garching, Germany.\\
\supit{h}NOVA,PO Box 9513, 2300 RA Leiden, Netherlands.
}

\authorinfo{Send correspondence to S. Kendrew, email: kendrew@strw.leidenuniv.nl; B. Brandl, email: brandl@strw.leidenuniv.nl}

\begin{document}
  \maketitle

\begin{abstract}
We present results of performance modelling for METIS, the Mid-infrared European Extremely Large Telescope Imager and Spectrograph. Designed by a consortium of NOVA (Netherlands), UK Astronomy Technology Centre (UK), MPIA Heidelberg (Germany), CEA Saclay (France) and KU Leuven (Belgium), METIS will cover the atmospheric windows in L, M and N-band and will offer imaging, medium-resolution slit spectroscopy (R$\sim$1000-3000) and high-resolution integral field spectroscopy (R$\sim$100,000). Our model uses a detailed set of input parameters for site characteristics and atmospheric profiles, optical design, thermal background and the most up-to-date IR detector specifications. We show that METIS will bring an orders-of-magnitude level improvement in sensitivity and resolution over current ground-based IR facilities, bringing mid-IR sensitivities to the micro-Jansky regime. As the only proposed E-ELT instrument to cover this entire spectral region, and the only mid-IR high-resolution integral field unit planned on the ground or in space, METIS will open up a huge discovery space in IR astronomy in the next decade.
\end{abstract}



\section{METIS: An overview}
\label{sec:metis}
METIS is the Mid-infrared E-ELT Imager and Spectrograph proposed for the 42-m European Extremely Large Telescope (E-ELT), which is expected to start operations in 2018 at Cerro Armazones (Chile). A Phase A study was carried out by a consortium of NOVA (comprising Leiden Observatory and Astron; Netherlands), the Max Planck Institute for Astronomy (MPIA; Germany), UK Astronomy Technology Centre (UKATC; UK), KU Leuven (Belgium) and CEA Saclay (France), in 2008-2009. Of the eight instruments under consideration for the Phase A stage, METIS is the only to cover the mid-infrared wavelength regime longward of 3 $\mu$m.

METIS is a multi-mode instrument covering the wavelength range of 3 to 14 $\mu$m, offering imaging, coronagraphy, medium-resolution slit spectroscopy and high-resolution integral field spectroscopy. An overall description of the instrument and its science goals are given by Brandl et al\cite{brandl10} in these proceedings (2010; paper [7735-86]). In addition, Lenzen et al~\cite{Lenzen2010} (2010; paper [7735-283]) show the instrument concept and trade-off solutions; Kroes et al~\cite{Kroes10} (2010; paper [7735-200]) describe the optomechanical design concept, including the cryogenic scheme; and Stuik et al~\cite{stuik10} (2010; paper [7736-127]) discuss the adaptive optics challenges and chosen solutions for METIS. All these papers are presented at these conferences, and we refer to them for a description of the instrument.

Modelling the performance of METIS served several aims: as well as providing a baseline reference for the science team, it allowed us to assess the impact of a number of environmental and observational parameters on the eventual performance of the instrument in its various modes of observation. The following sections present the method of calculation and an overview of the input parameters adopted and the resulting sensitivities. From the results we provide a comparison with current and next-generation facilities, and discuss the impact of a number of interesting parameters on METIS' projected performance.

\section{Performance calculations}

The sensitivity model developed for METIS builds on those constructed for other IR instruments of its kind. In particular, it references work by Swinyard et al (2004)~\cite{Swinyard2004} and the E-ELT exposure time calculators~\footnote{available online via http://www.eso.org/sci/facilities/eelt/science/drm/cases.html} developed by J. Liske for ESO's E-ELT Design Reference Mission~\cite{etc_spec, etc_im}. In fact, particular effort was made to ensure consistency with the latter. We describe here the method of calculation and an overview of the most important input parameters to the model. We focus here on the performance of three of METIS' observation modes: imaging, medium-resolution spectroscopy and high-resolution IFU spectroscopy; modes not modelled are polarimetric imaging and spectroscopy, and coronagraphy. Pantin et al (2010) ~\cite{pantin10} show the results of specific coronagraphic performance modelling with METIS in these proceedings (paper [7735-307]).

\subsection{Inputs and assumptions}

The input parameters to the sensitivity model for METIS come from a variety of sources. Specifications of the E-ELT telescope, site, and many observational and operational constraints were provided by ESO based on the latest available knowledge circa autumn 2009. Instrumental parameters are taken straight from the final design of METIS as of November 2009, manufacturer specifications where available, or extrapolations from existing instruments.

\subsubsection{Telescope}

\begin{table}
\centering
\begin{tabular}{lr}
\hline
Telescope parameter & Value\\
\hline
Primary mirror area & $\pi \times (21)^2 = 1385.4\;m^2$\\
Central obscuration & 9.2\% of primary area = $0.092 \times 1385.4 = 127.5\;m^2$\\
Total effective collecting area (A) &  $1385.4-127.5 m^2= 12.6 \times 10^6 \; cm^2$\\
Telescope mirror reflectivity & 98\%\\
Telescope transmission ($\tau_{tel}$) & $0.98^5=0.903$\\
\hline
\end{tabular}
\caption{Basic telescope parameters}\label{tab:telescope}
\end{table}

Table~\ref{tab:telescope} shows input parameters for the telescope, as provided to the community by ESO, and used as input to the calculations.

\subsubsection{Thermal background}\label{inp:bgr}

The thermal background is of crucial importance in the performance of infrared instruments, both in space and on the ground. Ground-based instruments are however particularly affected by the high level of background radiation from the atmosphere and the telescope. Dealing with these contributions appropriately is key to the accurate calculation of METIS' performance; particularly given the desire to operate METIS in the background-limited regime (BLIP) in all bands and observation modes. Three factors contribute to the thermal background seen at the METIS detector:

\begin{itemize}
\item{Atmosphere}
\item{Telescope}
\item{Instrument}
\end{itemize}

For the instrumental background we model only the background radiation from the entrance window. This window will consist of an approximately 1.2 cm plate of ZnSe, whose emissivity rises strongly between 12 and 20 $\mu$m. As the window will effectively be at ambient temperature, this can add a non-negligible background component in the N band. Our models show however that the contribution of the entrance window is small, thanks to transmission properties of dichroics, which filter out most radiation longward of 20 $\mu$m.

The atmosphere acts as a strong radiator at IR wavelengths; the emissivity profile depends on the vertical profiles of temperature, pressure and molecular abundances (e.g. $H_{2}O$, $CO$, $CH_{4}$), and the telescope's altitude above sea level. In these calculations we use a Paranal model atmosphere that provides an approximation for the observational conditions at nearby Armazones. In addition, the performance of METIS was calculated for a typical 5000-m high and dry site. We do not consider any residual systematic flux from the background subtraction, only the photon noise of the background radiation itself. Table~\ref{tab:sites} shows the basic site parameters for Paranal and the high \& dry site; also included are the corresponding values for Armazones, as described by Sch\"{o}ck et al (2009)~\cite{Schock2009} and Ot\'{a}rola et al (2010)~\cite{Otarola2010}.

\begin{table}
\centering
\begin{tabular}[h]{lccc}
\hline
Site & Altitude (m) & Temperature (K) & pwv (mm)\\
\hline
Paranal & 2600 & 286. & 2.3\\
High \& dry & 5000 & 270. & 0.5\\
\emph{Armazones} & \emph{3064} & \emph{281.} & \emph{2.9}\\
\hline
\end{tabular}
\caption{Site characteristics used in generation of transmission and radiance profiles. Paranal and high \& dry values from the ESO DRM; Armazones values from Sch\"{o}ck et al (2009)~\cite{Schock2009} and Ot\'{a}rola et al (2010)~\cite{Otarola2010}.}\label{tab:sites}
\end{table}

\begin{figure}[htb]
\subfloat[]{\includegraphics[height=6cm]{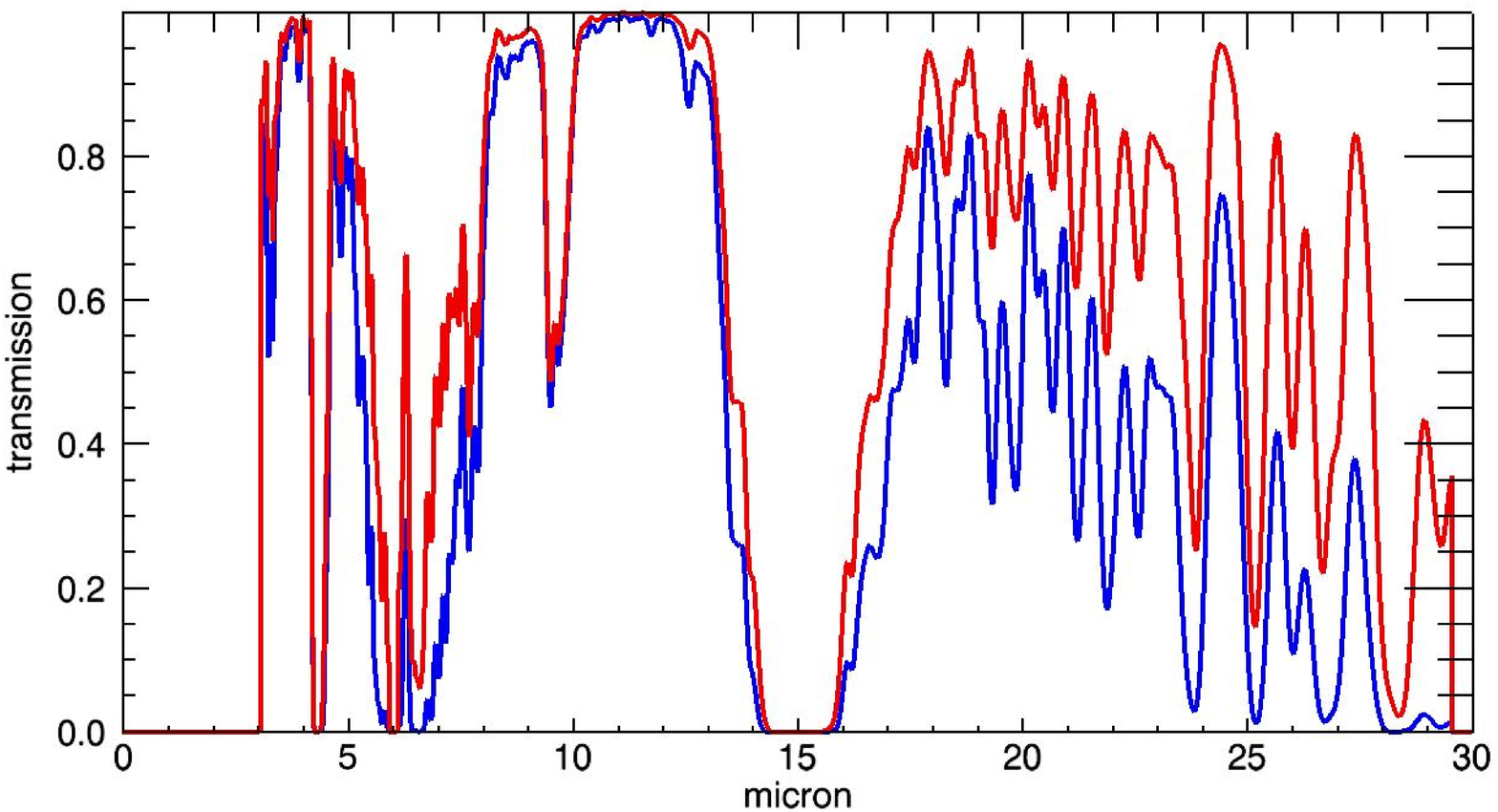}}
\subfloat[]{\includegraphics[height=6cm]{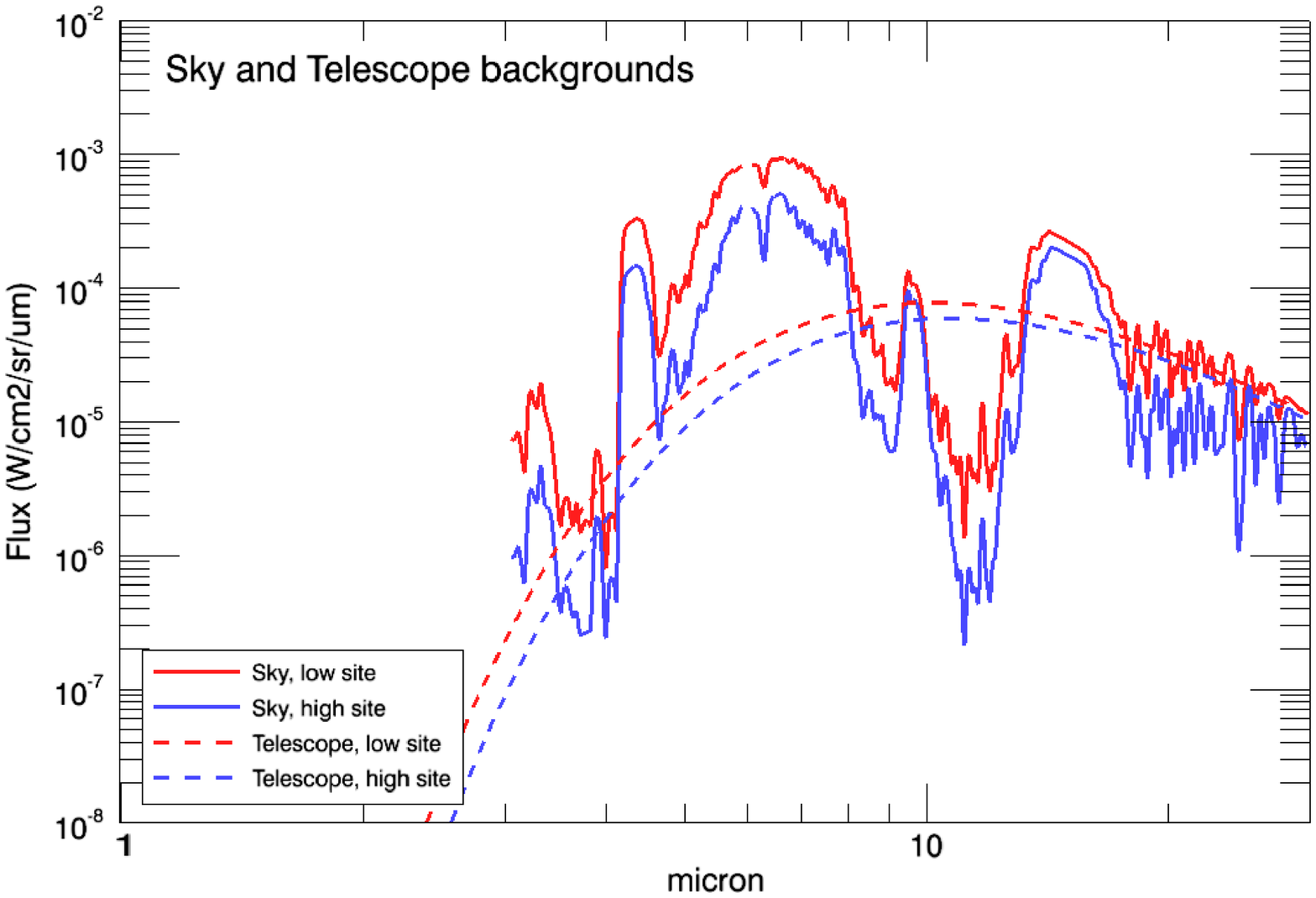}}
\caption{Comparison of site profiles used in the calculations. (a) Transmission; (b) Radiance from telescope (dashed lines) and sky (solid lines) for Paranal (red) and a high \& dry (blue) observing site.}\label{fig:sitecomp}
\end{figure}

Atmospheric transmission and radiance profiles were provided to the community by ESO, these were calculated using the publicly available Reference Forward Model (RFM)\footnote{http://www.atm.ox.ac.uk/RFM/atm/} with the HITRAN 2004 molecular line database~\cite{Rothman2005}, assuming a model tropical atmosphere and including the most common atmospheric molecules $H_{2}O$, $CO$, $CO_{2}$, $CH_{4}$, $O_{2}$, $O_{3}$ and $N_{2}O$. The profiles were produced with R$\sim$ 100000, and subsequently convolved to the appropriate resolutions for imaging and medium-resolution spectroscopy with METIS. The transmission profiles at the two considered sites are shown in Figure~\ref{fig:sitecomp}.

The telescope itself radiates as a grey body, at a temperature assumed equal to the ambient temperature appropriate to the site (see Table~\ref{tab:sites}). As a baseline we assumed a lower limit of 10\% was assumed for the telescope emissivity, and in section~\ref{sec:emissivity} the effect of a higher emissivity is investigated.


\subsubsection{Detectors}\label{inp:detectors}

METIS will use two different types of science detectors. The short wavelength L and M bands images and spectra will be imaged onto 2k $\times$ 2k HgCdTe detectors. The N-band modes will employ a 1k $\times$ 1k Si:As Aquarius detector. Table~\ref{tab:detectors} shows the specifications for each type. Quantum efficiencies were adopted from manufacturer specifications.

\begin{table}
\centering
\begin{tabular}[hbt]{lcc}
\hline
 & LM band & N band\\
 \hline
 no. pixels & 2k x 2k & 1k x 1k \\
 pixel size ($\mu$m) & 18 & 30 \\
 dark current (e-/s) & 0.5 & 2000 \\
 read noise (e-/frame) & 20 & 1000 \\
 well depth & 120,000 & 1.4e+7 \\
 \hline
 \end{tabular}
 \caption{Detector characteristics (per pixel).}\label{tab:detectors}
 \end{table}

Given the well depth of the detector and the high thermal background flux from sky and telescope, particularly for broad-band observations, the detector integration time is calculated such that the counts do not exceed the half-well depth to avoid saturation or non-linearity effects. In high-resolution spectroscopy mode, where the background flux is heavily diluted, the integration time was calculated to ensure background noise limited, rather than read noise limited, performance.

Finally, a 20\% overhead factor was applied to the total on-source integration time, such that a 1-hour exposure represents $0.8\times3600s = 2880s$ of on-source integration time. We do not take into account any extra duty cycle related to the chopping and nodding; essentially we assume that both the chop and nod mechanisms will keep the source in the field. This is not always the case, therefore the duty cycle could be substantially lower in practice.

\subsubsection{Adaptive optics}
METIS will use adaptive optics (AO) to achieve diffraction-limited performance. The baseline mode of operation uses single-conjugate adaptive optics (SCAO), with a single natural guide star as reference source. Ensquared energies were calculated for METIS in SCAO mode using the analytical code PAOLA~\cite{Jolissaint2006}. The simulations assumed seeing of 0.65'' at a zenith angle of 30$^{\circ}$ (giving a line-of-sight seeing of 0.71''), an outer scale of 25 m and a guide star V-band magnitude of $\leq$ 10. We assume the same AO performance at the low and high sites.

A laser tomography AO facility is expected to be added to the E-ELT suite of instruments in due course after first light~\cite{Fusco2010}, and this will help METIS reach maximal sky coverage. Using ensquared energy values produced by the LTAO unit study team, the performance of METIS in LTAO mode could also be investigated in these calculations.

Assuming the source photon noise is negligible $sensitivity \propto 1/ee$ (see equation~\ref{eq:nobj} in the next section). This automatically gives an estimate of the difference in sensitivity resulting from the switch to LTAO mode: the sensitivity of METIS with LTAO will be around 15-20\% lower in L and M band than with SCAO, and up to 40\% in N band. In addition to giving a reduced in ensquared energy, the LTAO mode is also likely to be a less efficient mode of observation, i.e. with larger overheads, to the simpler SCAO configuration. Given the relatively early stage of development of the LTAO module and the lack of on-telescope experience with advanced AO architectures such as LTAO, these numbers should be considered as preliminary estimates.

\subsection{Calculation method}\label{sec:calcs}

For mid-infrared ground-based astronomy, the signal to noise (S/N) of an observation is given by:

\begin{equation}\label{eq:sn}
S/N=\frac{\sqrt{frames}\times n_{obj}}{\sqrt{n_{obj}+n_{sky}+n_{pix}\times (rd)^{2}+n_{pix}\times (dark) \times (DIT)}}
\end{equation}
where:
\begin{description}
\item frames = number of frames per exposure
\item DIT = detector integration time per frame [s/frame]
\item $n_{obj}$ = detector counts from the object in the S/N reference area [e-/frame]
\item $n_{sky}$ = detector counts from the sky background in the S/N reference area [e-/frame]
\item $n_{pix}$ = number of pixels in the S/N reference area, determined by sampling properties of the observing mode
\item rd = detector read noise per pixel [e-/frame]
\item dark = detector dark current per pixel [e-/s]
\end{description}

Ignoring the source photon noise ($\sqrt{n_{obj}}$) against the strong thermal background, the sensitivity can be calculated as the minimum detectable signal at a S/N of 10 in a 1 hour exposure:

\begin{equation}\label{eq:minsig_cts}
minsig=\frac{S/N \times \sqrt{n_{sky}+n_{pix}\times (rd)^{2}+n_{pix}\times (dark) \times (DIT)}}{\sqrt{frames}}\;\;\;\;\;[e\!-\!]
\end{equation}

which can be converted back into flux units. We can now expand, first on some basic inputs related to the telescope and detectors, then on the calculation of key parameters in eqns~\ref{eq:sn} and~\ref{eq:minsig_cts}.

We define a factor $conv$, which converts the incident photon flux at the telescope to detected photo-electrons:

\begin{equation}\label{eq:conv}
conv=\frac{A \times \tau_{tel} \times \tau_{ins} \times (dqe) \times (pcg) \times \Delta \lambda}{E_{\gamma}}\;\;\;\;\;[e\!-\!/\!s\; per\; W\!/\!cm^{2}\!/\!\mu m]
\end{equation}

where

\begin{trivlist}
\item $A =$ photon collecting area [$cm^2$]
\item $\tau_{ins} =$ instrument throughput
\item $\tau_{tel} =$ telescope transmission
\item $dqe =$ detector quantum efficiency
\item $pcg =$ photo-conductive gain = 1 $e\!-/photon$
\item $\Delta\lambda =$ filter width or width of resolution element [$\mu$m].
\end{trivlist}

\subsubsection{Counts from the astronomical source ($n_{obj}$)}\label{eqns:nobj}

For an unresolved point source, the number of electron counts at the detector from the astronomical source, within the defined S/N reference area, is given by:

\begin{equation}\label{eq:nobj}
n_{obj}=F'_{obj} \times (ee) \times (conv) \times (DIT)\;\;\;\;\;[e\!-\!/\!frame]
\end{equation}

where $conv$ as defined above, and:

\begin{trivlist}
\item $F'_{obj}$ = the flux from the source arriving \emph{at the telescope},  given by $\tau_{atm} \times F_{obj}$, where $\tau_{atm}$ is the atmospheric transmission and $F_{obj}$ the source flux at the top of the atmosphere [$W\!/\!cm^{2}\!/\!\mu m$]
\item $ee$ = the encircled energy, i.e. the fraction of energy from the source radiation within the S/N reference area. This quantity is obtained from lookup tables.
\end{trivlist}

\subsubsection{Counts from the background ($n_{sky}$)}\label{eqns:nsky}

Considering only the photon noise, the detector counts from the thermal background within the S/N reference area per exposure can be calculated from:

\begin{equation}\label{eq:nsky}
n_{sky}=F_{sky} \times \Omega \times (conv) \times (DIT)\;\;\;\;\;[e\!-\!/\!frame]
\end{equation}

where:

\begin{trivlist}
\item $F_{sky}$ = the flux from the thermal background $[W\!/m^2/\mu m/arcsec^2]$, defined as sky emissivity $\times \tau_{tel} + $ telescope background + entrance window background. The sky, telescope and window emissivity models are described in section~\ref{inp:bgr}.
\item $\Omega$ = size of the S/N reference area $[arcsec^2]$. The S/N reference area is taken to be square in geometry, with $\sqrt{\Omega}$ given by the optical design of the given instrument mode.
\end{trivlist}

\section{Projected sensitivity of METIS}

The calculations outlined above were applied to METIS' three main observation modes: imaging, medium-resolution spectroscopy and high-resolution spectroscopy. We discuss here some specifics for each of these modes and present sensitivity results. All sensitivities are calculated for unresolved point sources at a S/N of 10 in 1 hour, and include the 20\% overhead factor described in section~\ref{inp:detectors}. The following sections show calculation results for each of METIS' main observing modes. We do not model the performance of METIS in coronagraphic imaging mode, a dedicated paper by Pantin et al (2010; paper [7735-307])~\cite{pantin10} in these proceedings describes this in more detail. As part of the phase-A study the option of an N-band channel to the high-resolution spectrograph was also investigated, with R$\sim$50000. This mode was not included in the baseline design but we present here the performance estimates for such a mode to demonstrate its scientific potential~\cite{brandl10}.

\subsection{Imaging}

The METIS imager covers a field of view of 18"$\times$18", with the beam split between two modules, one covering L and M bands, the other for N band, by a dichroic. The sampling is optimised (Nyquist sampled) at 3.5 $\mu$m and 7 $\mu$m for the short- and long-wavelength modules, respectively. This results in pixel scales of 8.6 mas/px for L/M band, and 17.2 mas/px for N band. The sensitivity was calculated over the square area encompassing the diffraction limited core of the PSF at the centre wavelength of the chosen filter - typically 3$\times$3 or 4$\times$4 pixels.

While the suite of METIS filters has not been definitively determined, we calculated sensitivities for typical broadband infrared filters, as used in today's instruments such as ISAAC and VISIR. Filter details per band are shown in Table~\ref{tab:filters}. Assuming 99\% reflectivity of the internal instrument optics and typical dichroic transmission properties, the throughput of the imager module (including the entrance window, internal optics and DQE, excluding atmopshere and telescope) was calculated to be $\sim$60\% in all three bands.

Results of the point source sensitivity calculations as described above are shown in Table~\ref{tab:imsens}, indicating a sub-microJansky performance in L-band for typical observing conditions at a Paranal-like site. Performance at a high \& dry site is shown for comparison.

\begin{table}
\centering
\begin{tabular}{lcc}
\hline
Band & $\lambda_c$ ($\mu m$)& $\Delta\lambda$ ($\mu m$)\\
\hline
L & 3.78 & 0.58\\
M & 4.66 & 0.10\\
N & 10.7 & 1.40\\
\hline
\end{tabular}
\caption{METIS imager filter definitions for the sensitivity calculations.}\label{tab:filters}
\end{table}
\bigskip
\bigskip

\begin{table}
\centering
\begin{tabular}{lccc}
\hline
Band & $\lambda_c (\mu m)$ & \multicolumn{2}{c}{Sensitivity, point ($\mu$Jy)}\\
\hline
 & & Low site & High \& dry site\\
\hline
L & 3.78 & 0.6 & 0.4  \\
M & 4.66 & 7.0 & 4.2 \\
N & 10.7 & 27.5 & 23.0 \\
\hline
\end{tabular}
\caption{METIS imaging sensitivities for L, M and N bands at low and high sites using SCAO. The point sensitivities were calculated for an unresolved point source observed for 1 hour to S/N of 10.}\label{tab:imsens}
\end{table}

\subsection{Medium-resolution spectroscopy}

The medium-resolution spectroscopy mode in METIS is achieved by inserting a slit and grism into the optical path of the imager. Its optical design parameters are therefore very similar to the imager mode described above. Spatial sampling properties for the medium-resolution spectroscopy mode are the same as for the imager. The slit width is set to $2\lambda_{slit}/D$ at the shortest wavelengths in the band, i.e. $\lambda_{slit} =$ 3 $\mu m$ and 7 $\mu m$ for LM and N bands, respectively. We assume that the spectral resolution element is sampled by 2 pixels.

For this mode, the sensitivity was calculated over the rectangular area covered by the slit width in one dimension, and the full width at half maximum of the PSF at the chosen reference wavelength in the other. This rectangular geometry causes an overestimation of the ensquared energy of around 5-10\%, and hence the sensitivity may be slightly worse than quoted by that amount.

A number of grisms are proposed for the low- and medium-resolution spectroscopy mode in METIS. For the sensitivity calculation we reference the two lowest-resolution grisms per band, each of which give one-shot full-band coverage for LM and N bands, respectively. The grism details are listed in Table~\ref{tab:grisms}.

\begin{table}
\centering\begin{tabular}{cccc}
\hline
Band & $\lambda_{min}$ ($\mu m$) & $\lambda_{max}$ ($\mu m$) & 2 px resolving power\\
\hline
LM & 2.8 & 5.2 & 1700\\
N & 8.0 & 14.0 & 940\\
\hline
\end{tabular}
\caption{Grisms for the medium-resolution spectroscopy mode for METIS.}\label{tab:grisms}
\end{table}

Typical throughputs in medium-resolution spectroscopy mode are 40-60\% for L and M bands, and $\sim$60\% in N band. The wide band coverage makes the spectrograph throughput sensitive to artefacts in the transmission profile of the dichroics, which are challenging to manufacture for IR wavelengths and subject to ongoing research\cite{Hawkins2008, Hawkins2008a}. The profiles used in these calculations are based on current samples produced for the JWST instrument MIRI.

Results of the continuum and line sensitivity calculations for unresolved point sources with the medium-resolution spectrograph are shown in Table~\ref{tab:mressens}. Reference wavelengths were chosen in each band at the location of interesting spectral markers from the METIS science case: the 4.05$\mu$m Br$\alpha$ line in L band, 4.7$\mu$m CO lines in M band, and 12.8$\mu$m [NeII] line in N band.

\begin{table}
\centering
\begin{tabular}{lccccc}
\hline
Band & $\lambda_c (\mu m)$ & \multicolumn{2}{c}{Low site} & \multicolumn{2}{c}{High site} \\
\hline
 & & Cont ($\mu$Jy) & Line ($\times 10^{-19} W/m^2$) & Cont ($\mu$Jy) & Line ($\times 10^{-19} W/m^2$)\\
\hline
L & 4.05 & 15 & 0.06 & 10 & 0.04  \\
M & 4.7 & 46 & 0.17 & 29 & 0.11 \\
N & 12.8 & 507 & 1.26 & 429 & 1.1 \\
\hline
\end{tabular}
\caption{METIS medium-resolution spectroscopic sensitivities for L, M and N bands at low and high sites. Calculated for an unresolved point and line sources observed for 1 hour to S/N of 10.}\label{tab:mressens}
\end{table}

\subsection{High-resolution integral field spectroscopy}

The high-resolution spectrograph in METIS in its baseline design consists of a single channel, separate from the imager and medium-resolution spectrograph optics, covering the L and M bands. An optional N-band channel with R$\sim$50000 was also investigated, eventually not adopted in METIS' baseline design. The spectrograph has an integral field design using an image slicer to divide the field spatially. In the spatial direction, the slice width is optimised to give critical (Nyquist) sampling at $\lambda_{spat} =$ 3.7 $\mu m$ and 9.0 $\mu m$ for LM and N bands respectively, i.e. the number of spatial pixels over which the sensitivity is calculated $pix_{spat}$ is given by $2 \times (\lambda/\lambda_{spat})$.

Spectrally the resolving power is designed to be 100000, optimised at 4.7~$\mu m$ for LM bands, and 50000, optimised at 12.8 $\mu$m for N band. At these locations an unresolved line will be sampled by 2.5 pixels (an anamorphism factor of 1.25 is introduced by the gratings). Although the variation of spectral sampling is affected both by the changing resolution element and the wavelength dependence of the dispersion function, we assume 2.5 pixel spectral sampling over the whole band, i.e. $pix_{spec}=2.5$.
The anamorphism essentially gives each pixel a rectangular, rather than square, field of view. The as-designed sampling in LM band is 9.0 mas/px spatially and 7.2 mas/px spectrally. The instantaneous wavelength coverage of the IFU spectrograph is  $\sim$60 nm. For the optional N-band channel, the as-designed pixel sampling is 22 mas/px spatially and 17.6 mas/px spectrally, with a wavelength coverage of $\sim$120 nm.

The spectrograph module contains significantly more optical components than the imager and its throughput is therefore lower (25-35\%).

The IFU sensitivities were calculated for unresolved lines from point sources, detected to S/N of 10 in 1 hour; results are shown in Table~\ref{tab:ifusens}. Given the high dilution of the sky and telescope background by the gratings, the risk of saturation is much lower than for broadband observations. The detector read noise, however, is substantial (see Table~\ref{tab:detectors}). The detector integration times per frame ($DIT$) were therefore calculated such that the noise is dominated by background noise, rather than read noise from the detector. At a Paranal-like site this requirement leads to typical integration times of $\sim$35s in L band, $\sim$10s in M band, and $\sim$60s in N band.

\begin{table}
\centering
\begin{tabular}{lccc}
\hline
Band & $\lambda_c (\mu m)$ & \multicolumn{2}{c}{Sensitivity ($\times 10^{-19} W/m^2$)}\\
\hline
 & & Low site & High \& dry site\\
\hline
L & 4.05 & 0.009 & 0.006\\
M & 4.7 & 0.026 & 0.017 \\
N & 12.8 &  0.133 & 0.116 \\
\hline
\end{tabular}
\caption{METIS high-resolution spectroscopic line sensitivities for L, M and N bands, at low and high sites, for unresolved point sources. Calculated for an unresolved line observed for 1 hour detected at S/N of 10.}\label{tab:ifusens}
\end{table}

\subsection{Comparison with current and future facilities}

\begin{figure}[h]
\centering
\includegraphics{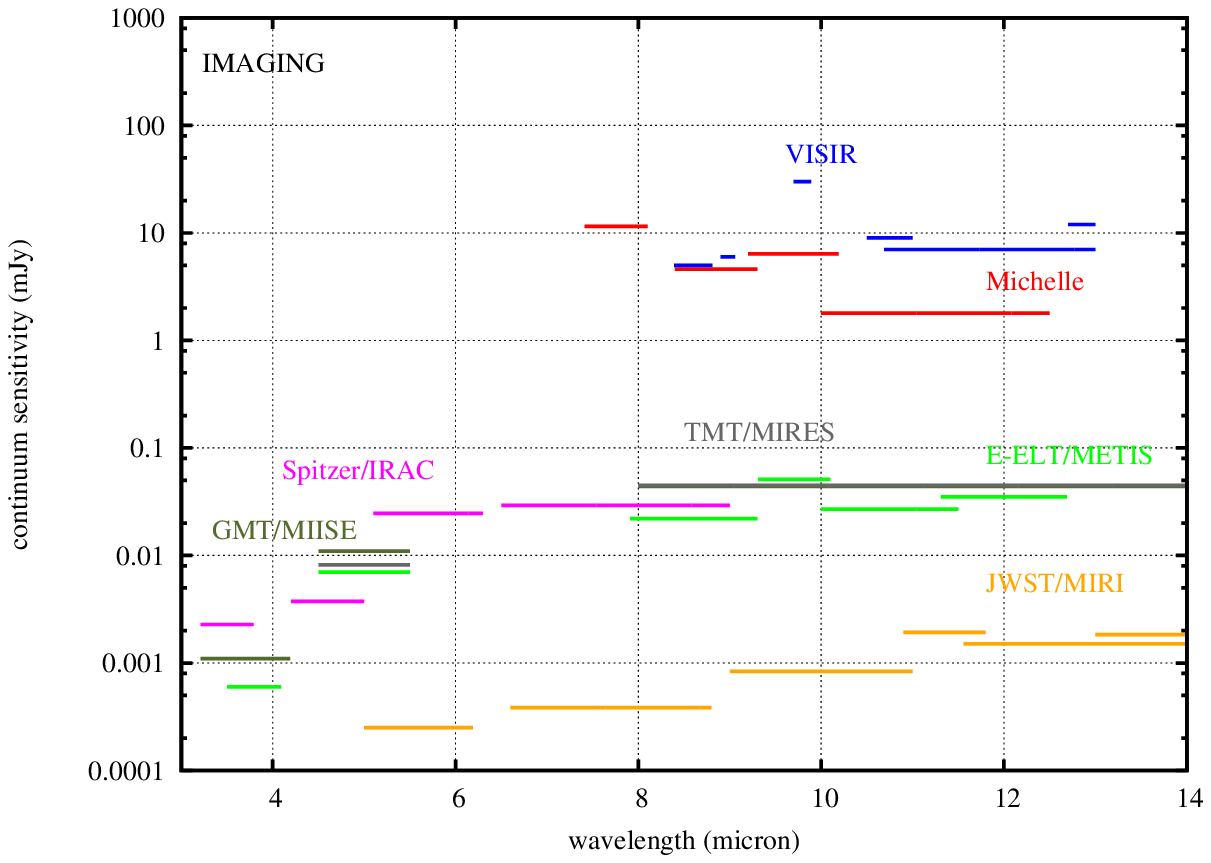}
\caption{Imaging performance of METIS compared with a number of current and contemporary facilities.}\label{fig:senscomp_im}
\includegraphics{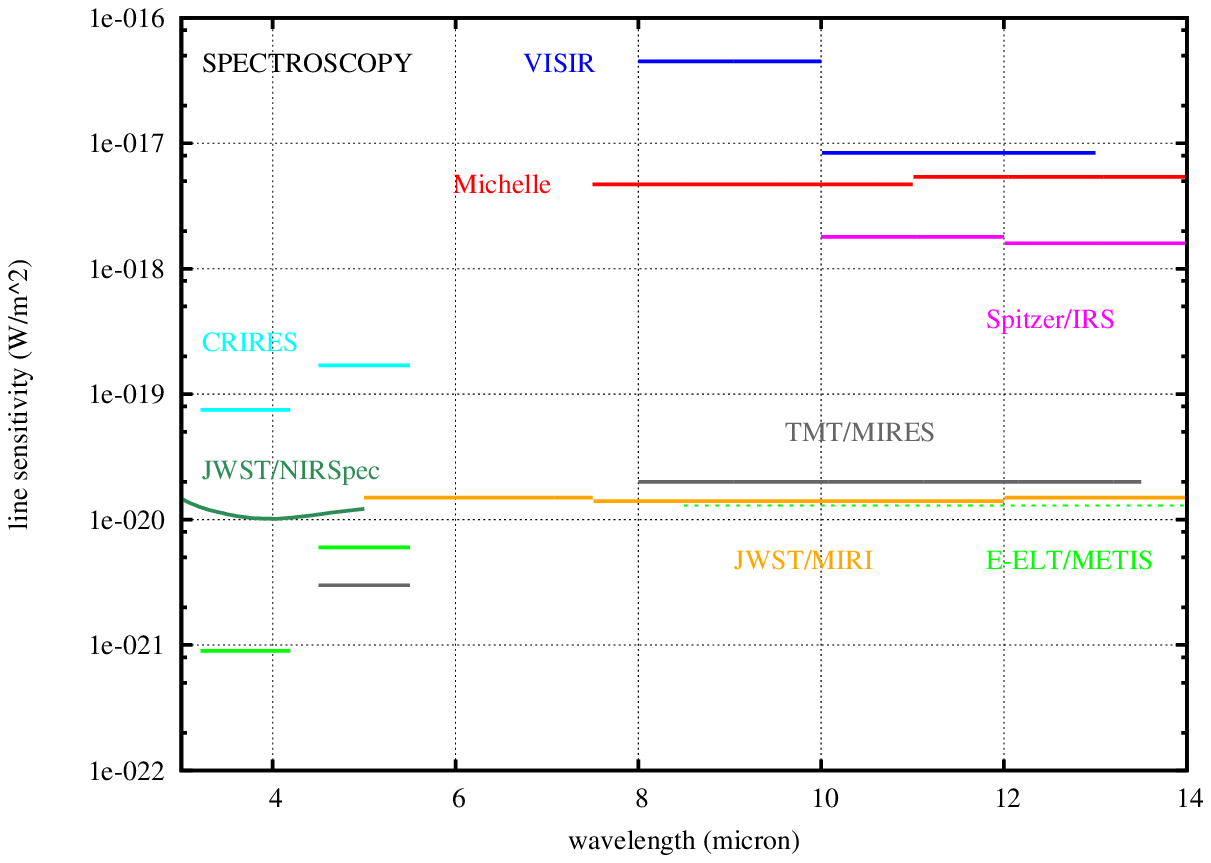}
\caption{Spectroscopic performance of METIS compared with a number of current and contemporary facilities.}\label{fig:senscomp_spec}
\end{figure}

The likely sensitivities of METIS, as presented here, were compared against the performance of other facilities operating in the mid-infrared wavelength range - either at present, or planned to come online in the next decade. Given the difficulties of observing in the mid-IR from the ground, a comparison with current and future space-based facilities, notably Spitzer and JWST-MIRI, is particularly interesting. The sensitivity values quoted in this section were all scaled to the same benchmark of a detection of an unresolved point source to 10-$\sigma$ in a 1-hour observation, using the scaling relations resulting from equation~\ref{eq:sn}. The spectroscopic sensitivities should be observed with care, given the different resolving powers offered by the various instruments. The comparative plots are shown in Figures~\ref{fig:senscomp_im} and~\ref{fig:senscomp_spec}, for imaging and spectroscopy, respectively.

For imaging, we compare the performance of METIS at a Paranal-like site to that of the current 8-m class instruments VISIR~\cite{lagage04} (VLT) and Michelle~\cite{BrysonIanR.1994} (Gemini-N), obtained from publicly available limiting magnitude or sensitivity values in the instrument manuals. In addition, the plot shows the performance of IRAC~\cite{Fazio2004} on board the Spitzer Space Telescope for high-background observations, as given in the instrument handbook. The performance of three contemporary facilities is also plotted against METIS: the Giant Magellan Telescope's  MIISE~\cite{miise}; the Thirty Meter Telescope's MIRES; and MIRI on board JWST~\cite{Swinyard2004}. For MIRES no published sensitivities were found, and we adopted the METIS sensitivities at a high \& dry site scaled to a 30-m aperture.

Figure~\ref{fig:senscomp_im} shows how METIS' imaging performance is comparable though somewhat better than that of its contemporary instruments on GMT and TMT; in certain parts of the spectrum MIRES would benefit from the high \& dry location of the TMT. Only MIRI on JWST will substantially outperform METIS, offering continuous mid-IR wavelength coverage down to sub-microJansky levels, albeit at much lower spatial resolution. In this way MIRI and METIS provide excellent complementarity in the parameter space they are able to explore.

For spectroscopy, METIS' line sensitivities are compared against VISIR, Michelle, MIRES and MIRI, as above, as well as CRIRES~\cite{ulli_crires2004} on VLT; Spitzer's Infrared Spectrograph IRS~\cite{Houck2004} and NIRSpec for JWST. Of these instruments, CRIRES is the most comparable to METIS in terms of wavelength coverage and spectral resolving power; Figure~\ref{fig:senscomp_spec} shows how METIS will offer an improvement in line sensitivity of more than an order of magnitude. Furthermore, its IFU design will offer the only high-resolution 3D spectroscopic capability at these wavelengths in the next decade. The N-band spectroscopic sensitivity, which is not included in the final baseline design, is indicated with a dashed line.

\section{Discussion}

\subsection{The E-ELT site}

\begin{figure}
\centering
\includegraphics[width=15cm]{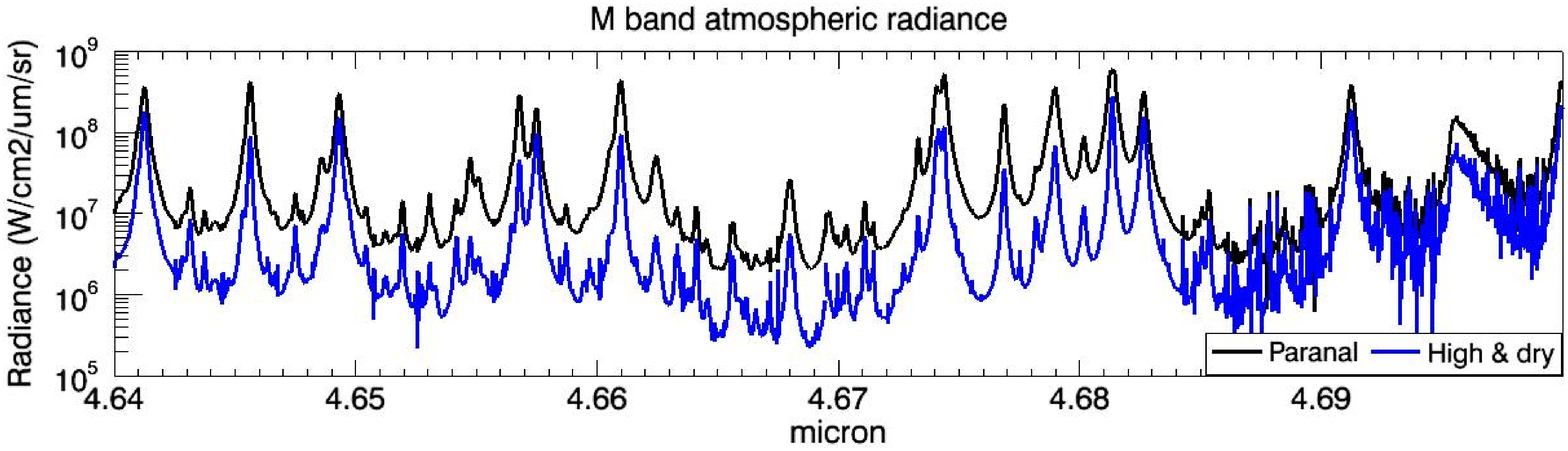}
\caption{Atmospheric radiance in M-band around the 4.7 $\mu$m CO emission band for low and high sites.}\label{fig:radiance_comp}
\end{figure}

An investigation into the instrumental performance cannot be decoupled from the atmospheric conditions at the telescope site. The atmospheric background dominates particularly the short edge of the L-band, the M-band and the long-wavelength edge of the N-band. Lower molecular abundances will improve the transparency of the window and lower the atmospheric radiance (e.g. in the CO-bands around 4.7 $\mu$m, see fig.~\ref{fig:radiance_comp}). In other regions, such as the short-wavelength part of the N-band, the telescope background dominates over the atmospheric radiance, and the benefit of a high site derives mainly from the lower ambient temperature.

For spectroscopy, a clear advantage is that even the strongest atmospheric absorption features are no longer saturated at very low humidity (typically above 5000m), allowing the observed spectra to be calibrated for telluric absorption with better accuracy (for a discussion of telluric line correction for METIS, see Uttenthaler et al (2010; paper [7735-269]) in these proceedings). During periods of very low atmospheric water vapour, the usually opaque 5-8 $\mu$m range can become accessible to ground-based astronomy.

\begin{figure}
\centering
\includegraphics[width=15cm]{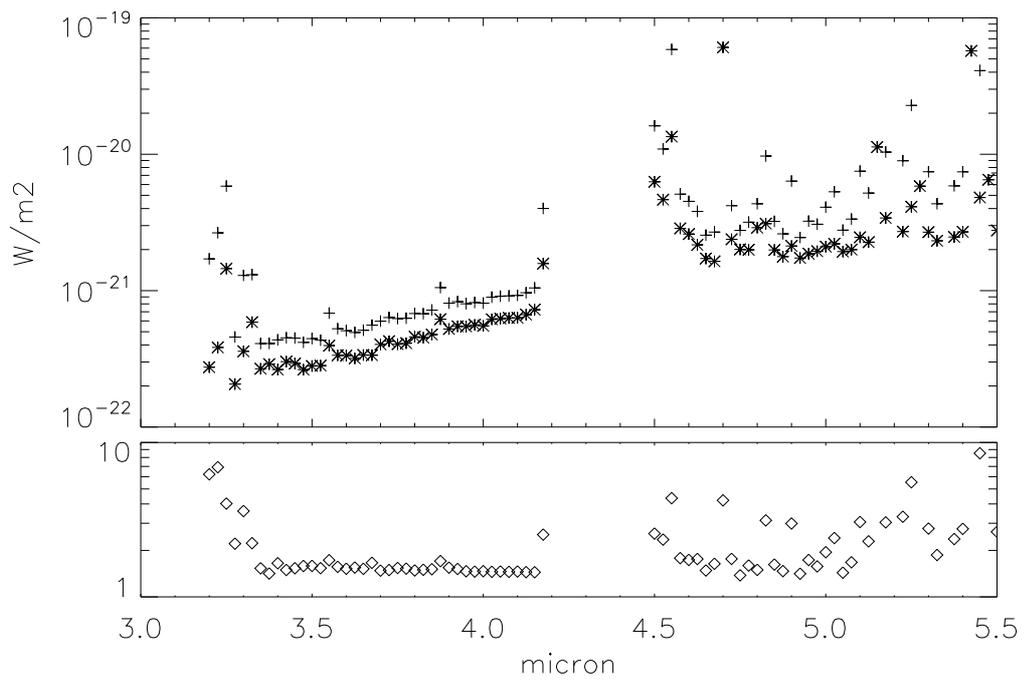}
\caption{Top panel -  METIS high-resolution lines sensitivities across L and M bands, for Paranal (+) and a high \& dry (*) site. Bottom panel - ratio low:high, showing clearly how the benefit of a high site is greater near the band edges.}\label{fig:ifusite}
\end{figure}

In general, the sensitivity values in Tables~\ref{tab:imsens}, \ref{tab:mressens} and \ref{tab:ifusens} indicate an improvement in L and M band of the order of 30\%, and around 15\% in N band at a high \& dry site. These values however generally cover band centres, where the atmospheric windows have maximal transparency. To investigate how the performance differs in the band edges, where atmospheric absorption is stronger, we calculate the high-resolution spectroscopic line sensitivities across the entire L and M bands for the two sites (see Figure~\ref{fig:ifusite}).

On inspection of the ratio between the sensitivities (bottom panel), we clearly see how the gain at the band edges is indeed higher - up to almost an order of magnitude in the range 3.2-3.4 $\mu$m (L band). Beyond 5 $\mu$m in M band the gain at the high site also rises above the band average.

In April 2010, the site for the E-ELT was announced to be Cerro Armazones~\cite{armazones_pr}, in the Chilean Andes close to the ESO site at Cerro Paranal. Basic site parameters, shown in Table~\ref{tab:sites}, suggest that the performance of METIS at Armazones will exceed that projected for Paranal. The site's higher altitude and lower temperature will result in a reduced thermal background and better transmission, however, the precipitable water vapour values found by Ot\'{a}rola et al (2010)~\cite{Otarola2010} indicate a somewhat wetter atmosphere. These quoted values may rely on different measurement techniques or instrumental calibrations and may not be suitable for a like-for-like comparison.

It is important to note that the relation between site characteristics and instrumental performance in the wavelength range covered by METIS goes beyond the bottom-level metrics we have described quantitatively here. Variability of parameters such as seeing and humidity affect the performance of the AO system, instrumental calibration and operations, and hence the final image quality. Detailed knowledge of the atmosphere at Cerro Armazones, both in basic meteorological terms (temperature, pressure, humidity) and turbulence profile over both vertical (turbulence layering and altitudes) and horizontal (outer scale) scales will be important in formulating an optimal calibration and operations strategy for METIS.

\subsection{Telescope emissivity}\label{sec:emissivity}

It is clear that the background from sky and telescope are an important, if not the most important, factor on the performance of any mid-IR ground-based instrument. The emissivity of the telescope is of particular concern, as its effect is twofold:

\begin{itemize}
\item higher emissivity increases the thermal background seen by the detector; and
\item higher emissivity implies a lower reflectivity, which leads to reduced throughput of the system.
\end{itemize}

This two-part effect of the telescope emissivity means the relationship between telescope emissivity and instrumental performance is not straightforward, depending on the telescope site (i.e. telescope temperature), the wavelength of observations and the filter bandwidth. In terms of the formalism shown in section~\ref{sec:calcs}, the emissivity will reduce the telescope transmission ($\tau_{tel}$) and increase telescope background contribution to $n_{sky}$.

The emissivity of the telescope is dominated by the emissivity of its largest optical surface, the primary mirror. The value of this parameter will be determined by the reflectivity of the mirror coatings, which may vary substantially between the segments; gaps in between the segments, which will act as strong (near-blackbody) infrared radiators; the presence of dust; and roll-off at the segment edges from the polishing process, which will cause extra light scattering and could potentially reflect warm sources near the telescope into the beam. In our calculations we assume the telescope radiates as a grey body at the ambient temperature of the site, i.e. irrespective of these properties, the telescope's background flux is lower at a high \& dry site. The telescope emissivity as specified by ESO is $\leq 10$\%.

\begin{figure}[ht]
\centering
\includegraphics[width=8cm]{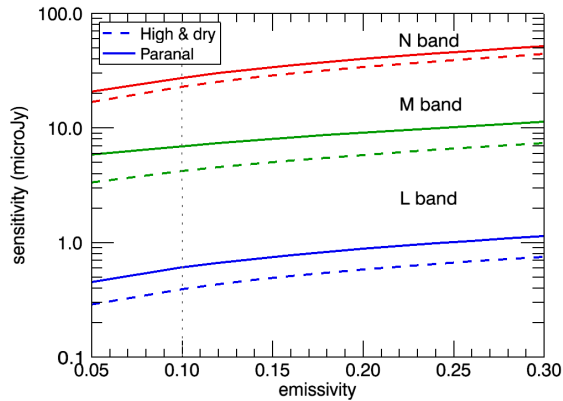}
\caption[Imaging sensitivities for METIS for telescope emissivities between 10 and 30 \%]{Imaging sensitivities for METIS for telescope emissivities between 5 and 30 \%, for L band, M band and N band. The Y-axis shows the absolute sensitivity value.}\label{fig:telemplots}
\end{figure}

To test the dependence on telescope emissivity of METIS' sensitivity, we recalculated the sensitivity for telescope emissivity values from 10 to 30 \% in imaging mode. The results, shown in Figure~\ref{fig:telemplots}, show the following:

\begin{itemize}
\item The deterioration of imaging sensitivity is quite dramatic, up to a factor 2 in L and N bands and a factor of 1.6 in M band. The precise value of the loss in sensitivity depends on band, filter and telescope site, however, a high telescope emissivity of 30\% will result in a loss of sensitivity of up to a factor of 1.5-2.0 at both a high and a low site.
\item The sensitivity requirements from METIS' ambitious proposed science programme~\cite{brandl10} will be met if the telescope emissivity remains $\leq 20$\% at a low site, and $\leq 25$\% at a high site. This is driven by the N band, where the infrared thermal background of both atmosphere and telescope is stronger than in L and M bands for a given temperature.
\end{itemize}

\section{Conclusions}

The calculations presented here were carried out during the 18-month Phase A study for METIS, in 2008-2009, and hence reflect the state of the telescope and instrument design as of late 2009. Based on our current knowledge and understanding, the sensitivity estimates convincingly demonstrate that METIS will be a formidable addition to the suite of ELT-class instruments worldwide, and moreover it is the only proposed first-light instrument on any ELT covering the full thermal and mid-IR ranges with imaging and spectroscopy. METIS will offer an order of magnitude improvement in sensitivity over current ground-based instrument, at superior spatial resolution aided by adaptive optics. In addition, METIS will complement very well its main space-based counterpart, MIRI on JWST, by offering high-resolution spectroscopy and a much better spatial resolution, albeit at a lower sensitivity.

By studying a number of critical aspects, we can state that achieving these sensitivities with METIS hinges on a detailed knowledge of the local atmosphere at the E-ELT site (atmospheric turbulence, thermal background behaviour), as well as an understanding of how such environmental effects are likely to influence the image quality; this will be an important area of study for the instrument team. The telescope emissivity will also play an important role, and efforts to reduce this will be very beneficial for METIS.

The code used to carry out these calculations is publicly available for download at http://tinyurl.com/metis-sens.

\bibliography{METIS_spie10}
\bibliographystyle{spiebib}

\end{document}